\documentclass[fleqn,twoside]{article}
\usepackage{espcrc2}
\usepackage{epsfig}

\def\NPBP{{\em Nucl. Phys.} B (Proc. Sup.)}
\def\NPB{{\em Nucl. Phys.} B}

\def\PRL{\em Phys. Rev. Lett.}

\usepackage{graphicx}
\usepackage[figuresright]{rotating}

\newcommand{\AmS}{{\protect\the\textfont2
  A\kern-.1667em\lower.5ex\hbox{M}\kern-.125emS}}

\title{The QCD String Spectrum and Conformal Field Theory}

\author{K. Jimmy Juge\address{Fermi National Accelerator Laboratory, 
P.O.Box 500, Batavia, IL 60510, USA},
Julius Kuti\address{University of California at San Diego,
La Jolla, CA 92093, USA}
and	
Colin Morningstar\address{Carnegie Mellon University,
Pittsburgh, PA 15213, USA}}

\begin{document}

\begin{abstract}

The low energy excitation spectrum of the critical Wilson surface is discussed 
between the roughening transition and the continuum limit of lattice QCD. 
The fine structure of the spectrum is interpreted within the framework 
of two-dimensional conformal field theory.
\vspace{1pc}
\end{abstract}

\maketitle

\section{INTRODUCTION}

We believe that a deeper understanding of the string theory connection
with large Wilson surfaces will require a precise knowledge
of the surface excitation spectrum and the determination of 
the universality class of Wilson surface criticality in the continuum
limit of lattice QCD. This approach will also require a consistent
description of the conformal properties of the gapless Wilson surface 
excitation spectrum.
In this short progress report we summarize our ab initio on-lattice 
calculations (a more extended status report was
published recently\cite{JKM0}).
In collaboration with Mike Peardon, we have also  
studied the spectrum of a ``closed" flux loop 
across periodic slab geometry (Polyakov line) by choosing
appropriate boundary conditions and operators for selected excitations
{\em without} static sources\cite{MP}.

\section{QCD STRING FORMATION}

The first attempt at a comprehensive determination of the rich energy 
spectrum of the gluon excitations
\begin {figure}[t]
\vskip -0.1in
\hskip -0.3in
\epsfxsize=3.3in
\epsfbox{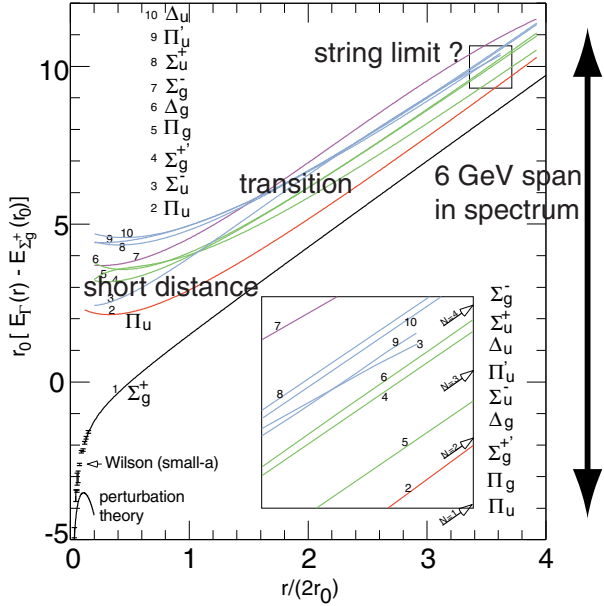}
\vskip -0.4in
\caption{Continuum limit extrapolations are shown for
the excitation energies where an arbitrary constant is removed
by subtraction. 
Color coding in postscript is added to the numerical labelling of
the excitations (N=0,black:1), (N=1,red:2), (N=2,green:4,5,6), 
(N=3,blue:3,8,9,10),
and (N=4,cyan:7). The five groups represent the expected quantum numbers
of a string in its ground state (N=0) and the first four excited states
(N=1,2,3,4).
The arrows in the inset represent the expected locations of the four lowest 
massless string excitations (N=1,2,3,4) which have to
be compared with the energy levels of our computer simulations.}
\label{fig:fig1}
\vskip -0.3in
\end{figure}
between static sources in the fundamental representation of ${\rm
SU(3)_c}$ in D=4 dimensions was reported earlier\cite{JKM1,JKM2}
for quark-antiquark separations r ranging from 0.1 fm to 4 fm. 
The extrapolation
of the full spectrum to the continuum limit
is summarized in Fig.~\ref{fig:fig1} with very different
characteristic behavior on three separate physical scales.
Nontrivial short distance physics
dominates for $r \leq 0.3~{\rm fermi}$. The transition region 
towards string formation is identified on the scale $0.5~{\rm fm}\leq
2.0~{\rm fm}$. String formation and the onset of
string-like ordering of the excitation energies
occurs in the range between 2 fm and 4 fm where
we reach the current limit of our technology.

To display the fine structure with some clarity, 
error bars are not shown in Fig.~\ref{fig:fig1}. Our earlier results are
compatible with extended new runs on our dedicated UP2000 Alpha cluster
which was built to increase the statistics more than an order of magnitude.
The notation and the origin of the quantum numbers used in 
the classification of the energy levels are explained in earlier
publications\cite{JKM1,JKM2}.
The physical scale is set by the Sommer ${\rm r_0}$ which, to a good
approximation, is ${\rm r_0=0.5~ fm}$. 

We also established that
the main features of string formation with three separate scales
is remarkably universal,
independent of the gauge groups SU(2) and SU(3),
and space-time dimensions D=3 and D=4: {\em Although
the level ordering is approximately string-like in all cases
at large separation,
there is a surprising and rather universal
fine structure in the spectrum with large 
displacements from the expected massless Goldstone levels.}

\section{CONFORMAL THEORY}

We believe that the fine structure of the Wilson surface spectrum can be
understood within the framework of two-dimensional conformal field theory.
This is illustrated first with the D=3 Z(2) gauge model. 
The Abelian subgroup Z(2) of SU(2) is expected
to play an important role in the microscopic mechanisms of quark
confinement suggesting that
Wilson surface physics of the D=3 Z(2) gauge spin model should
have qualitative and quantitative 
similarities with the theoretically more difficult
${\rm QCD_3 }$ case. 
In the critical region of the Z(2) model we have
a rather reasonable description of continuum string formation
based on the excitation spectrum of a semiclassical defect line 
(soliton) of the equivalent ${\rm \Phi^4}$ field theory.
The surface physics of the Z(2) gauge model is closely related to the BCSOS
model by universality argument and a duality transformation:
{\em their surface spectra should show universality.}

\subsection{BCSOS Surface Spectrum}

The body-centered solid-on-solid (BCSOS) model is obtained from the SOS
condition (accurate to a few percent around the roughening
transition) on the fluctuating interface in the body-centered
cubic Ising model\cite{BJ}.
This model can be mapped into the six-vertex formulation for which
the Bethe Ansatz equations are known\cite{EL}. It follows from
the Bethe Ansatz solution that the surface has a roughening
phase transition at 
${\rm T_R = J/(k_B\cdot 2ln2)}$ which is of the
Kosterlitz-Thouless type.
For ${\rm T < T_R}$ the interface is smooth
with a finite mass gap in its excitation spectrum. For ${\rm T \geq T_R}$ 
the mass gap vanishes and the surface exhibits a massless
excitation spectrum.

We determined the low energy part of the full surface spectrum from
direct diagonalization of the transfer matrix of the BCSOS model
and from the numerical solution of the Bethe Ansatz equations.
A periodic boundary condition was used, which corresponds to the spectrum
of a periodic Polyakov line in the Z(2) gauge model. With a flux of period
L we used exact diagonalization for ${\rm L\leq 18}$, and 
the Bethe Ansatz equations up to L=1024. 
The following picture emerges from the calculation for large L values
in the massless Kosterlitz-Thouless (KT) phase.
The ground state energy of the flux is given by
\begin{equation}
{\rm E_0(L)=\sigma_\infty\cdot L - \frac{\pi}{6L}c + o(1/L)~,}
\label{eq:bcsos1}
\end{equation}
where $\sigma_\infty$ is the string tension, c designates the conformal
charge, which is found to be c=1 to very  high accuracy.
The o(1/L) term designates the corrections to the leading 1/L behavior;
they decay faster than 1/L.
At the critical point of the roughening transition, the corrections
can decay very slowly, like ${\rm 1/(lnL^3\cdot L)}$ for the
ground state energy. Away from the
critical point, the corrections decay  faster than 1/L in power-like
fashion. {\em These finite size (or equivalently, finite
cut-off) effects in the fine structure of the spectrum are dominated
by the Sine-Gordon operator in conformal perturbation theory}\cite{JKM4}.

For each operator ${\rm O_\alpha}$ which creates states from the
vacuum (surface ground state) with quantum numbers $\alpha$,
there is a tower excitation spectrum,
\begin{equation}
{\rm E^\alpha_{j,j'}(L)=E_0(L) + \frac{2\pi}{L}(x_\alpha + j + j') + o(1/L)~,}
\label{eq:bcsos2}
\end{equation}
where the nonnegative integers j,j' label the conformal tower and
${\rm x_\alpha}$ is the anomalous dimension of the operator ${\rm O_\alpha}$.
The momentum of each excitation is given by
\begin{equation}
{\rm P^\alpha_{j,j'}(L)=\frac{2\pi}{L}(s_\alpha + j - j') ~,}
\label{eq:bcsos3}
\end{equation}
where ${\rm s_\alpha}$ is the spin of the operator ${\rm O_\alpha}$.
We find excitations of conformal towers built on the integer 
scaling exponents x=1,2 which are independent of the coupling  
and correspond to naively expected string excitations.
However, we also find scaling 
dimensions ${\rm x_\alpha}$ which continuously vary with the  
coupling J in the rough phase.
This sequence can be labelled by anomalous dimensions
\begin{equation}
{\rm x^G_{n,m} = \frac{n^2}{4\pi K} + \pi Km^2~,}
\label{eq:bcsos4}
\end{equation}
where n,m are non-negative integers and the constant K depends in a known 
way on the BCSOS coupling constant J. The physical interpretation of
the rather peculiar excitations of the rough gapless surface will be discussed
elsewhere\cite{JKM4}.  The surface spectrum is described by
a compactified conformal Gaussian field. The nontrivial 
part of the spectrum corresponds to field configuration 
with line defects which describe the screw dislocation pairs of the
fluctuating rough surface.

\section{D=3 QCD STRING THEORY}

If the $Q\overline Q$ color sources are 
located along one of the principal axes on the lattice in some
spatial direction, 
the Wilson surface at strong coupling is {\em smooth} 
in technical terms. This implies the existence of a mass gap 
in its excitation spectrum, as seen for example in the
strong coupling tests of our simulation technology.
As the coupling weakens, a roughening transition is expected in the surface at 
some finite gauge coupling ${\rm g=g_R}$ where the gap 
in the excitation spectrum 
vanishes with the characteristics of the Kosterlitz-Thouless
phase transition. 
The correlation length in the surface diverges at ${\rm g_R}$
and it is expected to remain
infinite for any value of the gauge coupling when ${\rm g \leq g_R}$.
At the roughening transition, the bulk behavior differs
from that of the continuum theory 
which is located in the vicinity of ${\rm g=0}$.
The low energy excitation spectrum of the Wilson surface 
for ${\rm g \leq g_R}$ and not far
from  ${\rm g_R}$, in the domain of the critical KT phase, should
be essentially identical to Eqs.~(\ref{eq:bcsos1}-\ref{eq:bcsos3}) 
of our BCSOS spectrum.

Now, is the critical Kosterlitz-Thouless picture around ${\rm g \leq g_R}$
identical to what we expect for the Wilson surface
in the low energy limit of continuum ${\rm QCD_3}$ string theory 
at ${\rm g=0}$? We guess that according to the most likely
scenario the Wilson surface remains massless 
throughout
the ${\rm 0 \leq g \leq g_R}$ region but its critical behavior will cross over
from the Kosterlitz-Thouless class of the conformal Gaussian
behavior into the universality class
of continuum QCD string theory whose precise description remains
the subject of our future investigations. The low energy effective
string Lagrangian will contain higher dimensional operators which
will signal the deviation from the Gaussian universality class.
These operators introduce a physical fine structure into the 
low energy spectrum. 
It will remain a challenge
to disentangle this physical fine structure from finite cut-off
effects in the surface which manifested themselves as finite size
corrections in the conformal spectrum at the roughening
transition.

\end{document}